# Gaze as a Supplementary Modality for Interacting with Ambient Intelligence Environments

Daniel Gepner[1], Jérôme Simonin[1], Noëlle Carbonell[1]

[1] LORIA, CNRS, INRIA & Nancy Université,
Campus Scientifique, BP 239, F54506, Vandoeuvre-lès-Nancy Cedex, France
{Daniel.Gepner, Jerome.Simonin, Noelle.Carbonell}@loria.fr

**Abstract.** We present our current research on the implementation of gaze as an efficient and usable pointing modality supplementary to speech, for interacting with augmented objects in our daily environment or large displays, especially immersive virtual reality environments, such as reality centres and caves. We are also addressing issues relating to the use of gaze as the main interaction input modality. We have designed and developed two operational user interfaces: one for providing motor-disabled users with easy gaze-based access to map applications and graphical software; the other for iteratively testing and improving the usability of gaze-contingent displays.

## 1 Context and motivation

Thanks to recent advances in gaze tracking techniques, it is now possible to consider the integration of gaze as an input modality into user interfaces. Eye trackers have become much less intrusive while maintaining satisfactory accuracy (0.5° visual angle). Users have not any more to keep their head and body perfectly still like in the nineties; they can move them with sufficient amplitude to feel comfortable while sitting at desktop PCs. In addition, remote eye trackers can be embedded in screen frames and coupled with standard video cameras; instant vision-based detection of the user's actual gaze direction, head position and orientation is used to correct raw point of gaze measurements and to adjust the angle of the eye tracking infrared camera(s). As for head-mounted eye trackers, which place no constraint on people's movements and mobility, they have become much less intrusive: their size and weight have decreased so much that they can be attached to spectacles. Finally, calibration has become much simpler and quicker. So, it is reasonable to expect that, in a few years, user gaze direction will be reliably detected and followed through "invisible" wearable devices. Vision-based gaze detection and tracking, a truly non intrusive approach, appears as an appealing solution in the long term. However, considerable research efforts and scientific advances are still needed to achieve appropriate accuracy despite steady significant advances [3], [10].

While technical feasibility is in the process of being achieved, usability issues are yet to be further investigated. Gaze is obviously a useful human-computer interaction

(HCI) modality for computer users with severe motor impairments [2], or for users who are engaged in demanding manual activities, such as pilots in cockpits; see the early study of gaze used to designate objects referred to in speech commands [8].

Contrastingly, the utility and usability of gaze as a supplementary input modality for able-bodied users has motivated much fewer studies and remains to be demonstrated. Gaze cannot of course supersede the mouse for pointing at, and selecting, graphical objects displayed on the screen of standard desktop PCs. As a pointing modality it can neither outmatch finger designation on a touch screen in efficiency and usability. However, its superiority to hand/arm gestures for remote pointing at large displays (e.g., "walls", reality centres or caves) is beyond doubt. Gaze pointing is faster, more precise and less tiring than gestures; hands are free and can be used to carry out other tasks, which may prove to be an invaluable advantage in some interaction contexts; in addition, it is more natural to use. Gaze, coupled with speech for expressing depth, can designate positions in a 3D virtual space more precisely and easily than 3D gestures; in particular, users have to mentally apply a homothetic transformation to their hand gestures and body moves to anticipate the actual span of these movements in a 3D virtual world, while planning them. Gaze pointing complemented with spoken phrases including action verbs is nevertheless inappropriate for manipulating virtual objects (e.g., on a Workbench screen), especially for shaping and deforming them. Therefore, gaze coupled with speech seems well suited to immersive virtual world exploration, but ill suited to virtual 3D object manipulation, as this form of input multimodality can be used for moving virtual objects but not for transforming them.

Research on the implementation of such concepts as 'Ambient Intelligence' and 'Ubiquitous/Pervasive Computing' is fast developing in the HCI scientific community. This evolution is likely to stimulate research on the integration of gaze into user interfaces as a remote pointing input modality. Ambient Intelligence (AmI), which subsumes Ubiquitous and Pervasive Computing, is a vision where humans are surrounded by computing and networking technology unobtrusively embedded in their surroundings. AmI lays emphasis on user friendliness, efficient and distributed services support, user empowerment, and advanced support to human interactions. This vision assumes a shift away from PCs to a variety of interactive "intelligent" agents which are unobtrusively embedded in our environment; see [5] for information on recent research advances on the implementation of AmI principles. Definitions of Ubiquitous and Pervasive Computing, [19], and [18], also put emphasis on "the integration of computation into our daily work practice to support our activities without being noticed" or even "visible"; which implies distributing interaction facilities in the environment and embedding them in everyday objects. So, future users of AmI and Ubiquitous/Pervasive Computing environments will be naturally induced to perform computer tasks while engaged in other non computing activities, at home or on the move. These parallel activities will sensibly reduce the perceptual and motor resources they will be able to invest in computing tasks. So, they will be confronted with similar computer access and interaction difficulties to those encountered by people with disabilities. Therefore, advances in research on Ambient Intelligence and progress in implementing this concept in software agents meant to support the general public in their daily activities are likely to significantly contribute to advancing the inclusion of people with disabilities in the fast developing Information Society, mainly by increasing computer access flexibility. They will also contribute to simplifying and unifying user-centred interface design, in-as-much as user interfaces meant to satisfy the needs and expectations of able-bodied users in a wide range of contexts of use will be easy to tailor to suit the specific needs of various communities of users with physical disabilities. The development of AmI in society will stimulate the progress of Universal Access.

These observations explain our interest for speech and gaze as a substitute multimodal command "language" for direct manipulation in some contexts of use, such as interaction with AmI artefacts or immersive virtual reality environments.

Our work on speech and gaze as a promising multimodal command "language" for interacting with augmented objects and large displays is presented in the following sections. We first specify our objectives. Then, the software tools that make up our experimental platform are described. Present research progress and outcomes are presented next. Future research directions are sketched in the conclusion.

## 2. Scientific objectives

Our main objective is to design and implement an appropriate speech- and gaze-based command language for interacting with large displays. Design requirements include the following major constraints. This multimodal language should be:
- robust, that is, tolerant to ambient noise and atypical or faulty utterances,
- flexible to changes of context of use and inter-individual expression differences;
- effective and efficient, and capable of outmatching existing interaction facilities;
- usable: in particular, easy to use and to learn (i.e., transparent).

To be usable, gaze pointing, as an input modality supplementary to speech, should be implemented in compliance with the following specific requirements:
- The role of gaze should be limited to solving deictic and underspecified reference phrases included in speech commands. For instance, the pronouns "that" and "there" in Bolt's famous example "Put that there." are deictics which refer to objects or positions on the user's screen [4].
- Interpretation of deictic phrases should be mainly inferred from the analysis of gaze fixation positions on the display. Minimum constraints should be placed on users' eye movements. Of course, inference of referents matching deictics from spontaneous eye movements would be an ideal solution; it represents our ultimate goal. But present knowledge and modelling of gaze activity during human-computer tasks is insufficient for interpreting spontaneous gaze movements accurately and reliably. Therefore, at least for the time being, controlled gaze pointing is mandatory; see section 4 for a presentation of our current work on this issue. Nevertheless, implementation of this constraint should interfere as little as possible with users' spontaneous visual exploration strategies. So, users might be asked to "look once at the objects and/or locations on the display referred to ambiguously in the current command"; yet, they should be free to choose when to look at referents (i.e., before, during and/or after the current verbal command), and for how long. In particular, dwells which have to last between 600 ms and 1 sec. to be reliably recognized in typing tasks on virtual keyboards [11] ought not to be required from users.

Research on speech- and gaze-based interaction is still in its early stages. Few studies have been published yet on the implementation and evaluation of gaze as a pointing modality: see, for instance, [17] and [6] for an assessment of pointing by gaze compared to standard pointing techniques, [13] for the use of this modality in pervasive computing environments, and [16] for its use in games. Multimodal interaction involving gaze has motivated fewer published studies; see, for instance, [20] on mutual disambiguation in gaze- and speech-based interaction.

One of the main difficulties is the well known "Midas touch" effect [9] which stems from the double origin of eye movement control: depending on the context and activity, gaze direction is either controlled by low-level perception mechanisms activated by the visual properties of objects in the environment (bottom-up control) or by intentional top-down processes. Fixations resulting from the latter processes are those which have the potential to yield information on the user's current intention. A way to avoid the Midas touch effect in gaze-based interaction is to avoid providing users with a visual feedback to their eye movements.

Our work on gaze as a supplementary input modality to speech is grounded on a main working hypothesis which is essential to obtain reliable and robust gaze and speech controlled user interfaces. We expect that users will behave as follows when using speech for controlling graphical software applications. There is some chance that they will glance at the graphical object on which the current oral command is designed to operate while planning the action they intend to perform on this object, and/or during the utterance of the command itself, and there is every chance that they will look at it once they have finished speaking, to check whether its execution by the system meets their expectations. If this assumption proves to be valid, it will be possible to distinguish between fixations driven by visual stimuli and those controlled by intents. Analysis of fixations occurring just before, somewhat after, and during the utterance of speech commands, is then likely to provide useful information on the object or action specified in the command.

## 3. Software tools

Numerous advanced software tools are necessary for studying the potential of gaze as a pointing modality. We have developed a software platform comprising a set of components with the following functionalities:

- Robust real time computation of gaze fixations using a dispersion-based algorithm. This component is necessary for implementing gaze as an input modality. See [14] for descriptions of the main fixation computation algorithms and comparisons of their performances. We use a head mounted eye tracker (ASL-501) at two sampling rates, 60 Hz or 240 Hz. Results of real time fixation computation are identical to those of the ASL Eyenal batch processing software.
- Recording of rich traces of user interactions with any Windows application software. Time stamped traces include:
    - system and user generated events;
    - screen displays (from 4 to 8 screen copies per second) and mouse phantom (or cursor) positions on the screen;
    - coordinates of the user's gaze points on the screen, and speech from both user and system; plug-ins can be added to process other modalities.

- Synchronised play back of multimodal interactions from files of recorded traces. Eye fixations appear superimposed on the displays; optional visual markers display qualitative information on pupil diameter, fixation duration and size (i.e., spatial dispersion of samples included in each fixation).
- Assistance to manual annotation of displayed traces so as to facilitate their qualitative and quantitative analysis. Main annotation facilities include:
  - Manual or semi-automatic segmentation and labelling of visual and audio traces so as to define meaningful sequences of interactions. For instance, replacing all occurrences of a phrase in a text by another phrase represents a meaningful sequence of interactions with a word processor.
  - Manual or automatic definition of interest zones on the displays so as to determine what graphical objects are most frequently looked at by users during the execution of a given task.
- Implementation of the Wizard of Oz paradigm, an efficient rapid prototyping technique for simulating functionalities of innovative user interfaces which aim at reproducing human expertises too costly or difficult to implement appropriately, such as natural language understanding. Traces of users' interactions are simultaneously recorded on the user PC and sent to the wizard PC via local network. It is possible to filter the data sent to the wizard to avoid overcrowding his/her screen. Wizard-to-user information is also conveyed through the local network. An application dependent plug-in meant to assist the wizard in simulation of the user interface can be added to the software component running on the wizard's PC.

A version of this distributed platform has been developed under Linux with the aim of obtaining appropriate observation tools for studying gaze- and speech-based interaction with 3D virtual environments. The overall functionalities of this platform are identical to those of the Windows tools described above. Main specific features originate from the chosen 3D development environment, a multi-agent system (oRis, ENIB, France) which represents each 3D object in the animated scene as an agent. This feature impacts on the platform functionalities as follows:

- Implementation of the Wizard of Oz technique: 3D animations generated by oRis run on both workstations; this strategy makes it possible to limit wizard-to-user data exchanges to the oRis commands necessary for simulating the execution of the user's multimodal requests, and thus, to easily obtain smooth evolution of the animation on the user's screen.
- The play back includes the names of the objects gazed at by the user during fixations, since agents can be labelled.

In addition, the play back displays pupil diameter evolution, phonetic and orthographic transcripts of the user's oral commands together with a graphical representation of the speech signal. Transcripts are either done by human experts or generated by acoustic-phonetic decoders and speech recognition systems.

Eye tracking contribution to HCI research is twofold. First, data on eye movements are most useful and are increasingly used for assessing the ergonomic quality of advanced graphical user interfaces and interactive visualisations of very large data sets. We have used the software platform presented here for eliciting the possible influence of display spatial layout on the efficiency and comfort of visual search [15].

Secondly, eye tracking data analysis is quite obviously an essential source of knowledge for implementing gaze as a pointing modality. The next section presents our scientific contribution to this emerging research area, and demonstrates the usefulness of the software tools presented above.

## 4. Gaze, a promising pointing modality

Our work on the implementation of gaze as an appropriate substitute modality for mouse pointing and selection in contexts where users interact with large displays, such as image walls or immersive virtual reality environments, is currently developing in three directions which are detailed next:
- prototyping of gaze controlled map-based applications,
- development of flexible gaze-contingent displays,
- design and development of "natural", gaze and speech command languages.

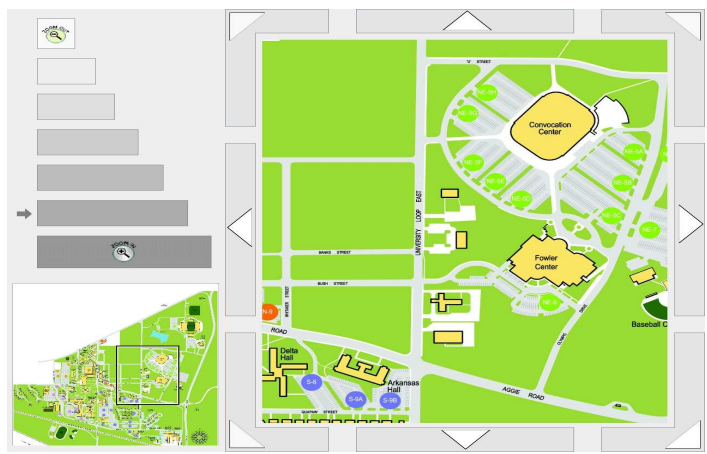

**Fig. 1.** Gaze-controlled user interface of the map-based application. Control buttons are in grey.

### 4.1. Gaze control of map-based applications

We have developed a prototype gaze controlled user interface for interacting with standard map applications. All the interaction facilities provided by applications on the Web, such as Mappy, can be activated using gaze sampled at a 60 Hz rate. This interface is presented in Figure 1. Looking at one of the four buttons surrounding the zoom-in window (on the right of the screen) moves the focus of the zoom to the right (left, top or bottom) by a fixed step the length of which can be adjusted by the user. The zoom-in factor can be specified using the seven buttons on the top left of the screen. We are currently considering implementation of continuous zoom-in.

Looking at a point on the overall map (at the bottom of the screen on the left) causes the current content of the zoom-in window to be replaced by a portion of the map the contours of which are represented on the overall map by a black rectangular line; the centre of this rectangle coincides with the averaged positions of the samples included in the fixation; its size varies according to the zoom-in factor. To avoid the Midas touch effect without constraining the user's eye movements, we have proceeded like this: (i) when a fixation on the overall map is detected, the system waits for about 170 ms from the start of the fixation; then, (ii) it displays a blue circle around the averaged x, y coordinates of the 10 first samples in the fixation to warn the user that it will change the content of the zoom-in window about 200 ms later; (iii) if the user goes on fixating the same position on the overall map for about 200 ms (i.e., 12 samples), the system replaces the content of the zoom-in window with the portion of the overall map surrounding the averaged coordinates of the 22 first samples in the fixation. These thresholds have been determined from the data collected in [15].

This prototype which has been developed to test the usability of gaze pointing as an input modality has been evaluated by a few users who have judged it intuitive and comfortable to use. Potential applications include accessible graphical software for motor disabled people, and interaction with large graphical displays, for instance, digital wall maps in lecture rooms.

### 4.2. Gaze-contingent displays

A gaze-contingent display is a display that responds to where an observer looks within it. The most informative details are generated at the current point of gaze, and the display resolution is degraded at the periphery of the foveal or parafoveal field of vision. Research on the implementation of gaze-contingent displays, an emerging area, is fast developing; see [7] for a review. Potential application areas include:
- Interaction with complex distant visualisations or animations in cases when image compression techniques are insufficient for achieving satisfactory display speed and reactivity; for instance, remote surgery through Internet is a promising application area.
- Interaction with virtual reality applications, especially in immersive environments such as reality centres or caves, where gaze represents a more efficient and usable pointing modality than hand or finger gestures.

We have developed a robust gaze-contingent prototype which functions as a gaze-controlled lens: the display is blurred outside a circle the centre of which coincides with the averaged coordinates of the 4 first gaze samples (i.e., 70 ms or so) included in the current fixation. See Figure 2. The display resolution is highest inside this circle the diameter of which corresponds to the standard size of the foveal field of vision, that is, 5° visual angle. This prototype is operational, and its accuracy has been successfully tested by 5 participants who used it for exploring complex realistic displays. However, participants complained that the "lens" followed their eye movements too slowly, which obliged them to sensibly reduce their natural pace of visual exploration in an irritating way. A 70 ms lag between fixation start and "lens" move is indeed too long, but it is necessary for achieving reliable fixation detection. The decision of restricting the high resolution area on the display to the span of the foveal vision field was motivated by the will to test the actual accuracy of our prototype. Obviously, this area should be extended so as to also include the parafoveal vision field, which might reduce users' frustrations significantly. The same motive explains the absence of smooth resolution degradation. The study described in [1] differs from ours in that the gaze-contingent lens creates a fisheye deformation, and the task amounts to target search and selection in a grid of 9x9 boxes.

We are currently investigating two research directions in parallel so as to overcome the main usability weakness of contingent displays, that is, their insufficient reactivity by: (i) increasing the sampling frequency rate, from 60 Hz to 240 Hz, with a view to shortening fixation detection time; (ii) achieving a satisfactory trade-off between accuracy and time delay by analysing saccade velocity variations to predict next fixation approximate landing. An approach combining both strategies will also be tested. Finally, to further improve the usability of gaze-contingent displays, we shall test whether smooth degradation of resolution is actually more comfortable visually than stepwise reduction.

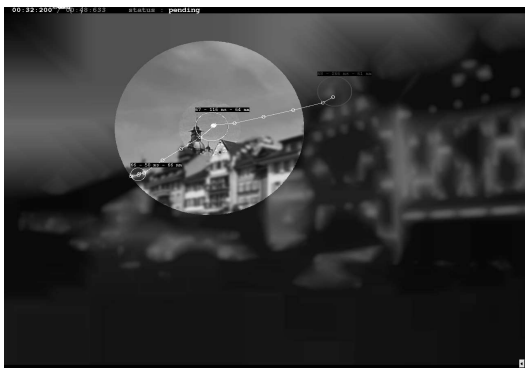 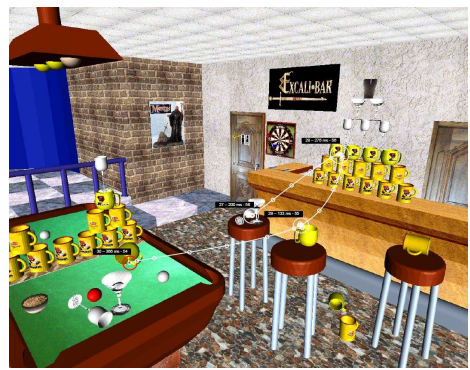

**Fig. 2.** Play back: gaze-controlled lens. Three fixations are visible; the one on the right is starting.

**Fig. 3.** Play back: Wizard of Oz simulation of gaze- and speech-based interaction.

### 4.3. Gaze, a supplementary modality to speech

We have collected a corpus of realistic speech and gaze interactions with various 3D environments to acquire the necessary knowledge for implementing a "natural" style of input multimodality, using the Wizard of Oz technique for interpreting users' eye movements and oral commands. See Figure 3. 5 participants performed 5 scenarios lasting 4 min. each. They carried out two types of interactive tasks, manipulation of virtual objects and interaction with a reactive game. They had to comply with a few expression constraints only:

- the vocabulary they could use was restricted to a predefined list of words;
- during the last scenario only, they had to designate each object they had to move using a deictic and "looking at" the object they wanted to move, then at the new position it should occupy; during the previous 4 scenarios they could move their eyes freely.

Participants' speech commands have been segmented into words and phones by an expert phonetician; time-stamped orthographic and phonetic transcripts of the speech corpus are now available. Gaze fixations are currently analysed to determine which ones correspond to moments when the user's visual attention is focused on the 3D object that the current[1] command aims to act upon. Then, various algorithms will be designed to filter controlled and spontaneous eye movements with a view to identifying fixations on task objects alluded to in the current command, especially by underspecified reference phrases, thus contributing to reduce the ambiguity and imprecision of speech commands. These algorithms will be trained, tested and evaluated on the corpus of multimodal interactions presented in this section.

---

[1] The oral command may be in progress or forthcoming or just completed.

## 5. Conclusion

We have presented our work on the implementation of gaze as an appropriate input modality for pointing at virtual interactive objects in various contexts of use: user interfaces with graphical applications for people with motor disabilities, gaze-contingent displays, especially on large screens and multimodal, speech- and gazed-based interaction with immersive virtual environments and augmented reality interfaces.
In the short term, our research on the possible contribution of gaze to enriching HCI will be focused on the following goals:
- Improve gaze tracking accuracy by taking account of pursuit movements the occurrence of which is very frequent during interaction with animated scenes.
- Increase the reactivity of gaze-contingent displays by experimenting with various approaches.
- Assess the effective contribution of spontaneous or loosely controlled gaze to disambiguating imprecise references in speech commands, interpreting deictic phrases, and improving speech recognition accuracy (see [12]).